\documentclass[pra,aps]{revtex4}
\usepackage{graphicx}
\usepackage{bm}

\newcommand{\ra}{\rangle}

\def\d{\dagger}
\def\>{\rangle}
\def\<{\langle}

\def\be{\begin{equation}}
\def\ee{\end{equation}}
\def\ba{\begin{eqnarray}}
\def\ea{\end{eqnarray}}
\def\bs{\begin{subequations}}
\def\es{\end{subequations}}
\def\diag{{\rm diag}}
\begin{document}

\title{Interference of two photons of different color}

\author{M. G. Raymer,$^1$ S. J. van Enk,$^1$ C. J. McKinstrie$^2$ and H. J. McGuinness$^1$}

\affiliation{$^1$Department of Physics and Oregon Center for Optics, University of Oregon, Eugene, OR 97403\\
$^2$Bell Laboratories, Alcatel--Lucent, Holmdel, NJ 07733}

\date{\today}

\begin{abstract}
We consider the interference of two photons with different colors in the context of a Hong-Ou-Mandel experiment, in which single photons enter each of the input ports of a beam splitter, and exit in the same, albeit undetermined, output port. Such interference is possible if one uses an active (energy-non-conserving) beam splitter. We find scenarios in which one ``red'' and one ``blue'' photon enter the beam splitter, and either two red or two blue photons exit, but never one of each color. We show how the precise form of the active beam-splitter transformation determines in what way the spectral degrees of freedom of the input photons should be related to each other for perfect destructive interference of the different-color components in the output. We discuss two examples of active beam splitters: one is a gedanken experiment involving a moving mirror and the other is a more realistic example involving four-wave mixing in an optical fiber.
\end{abstract}
\maketitle

\section{Introduction}

Bosons, in contrast to fermions, tend to occupy the same state when given the opportunity. This tendency arises from the fundamental quantum commutation relations of bosonic operators, and is responsible, for example, for the creation of laser light and the Bose-Einstein condensate (BEC). In the case of coherent matter waves, such as those in a BEC, stimulated scattering of bosonic atoms enhances the probability for the number of atoms in an already occupied atomic mode (in particular, the ground state of their external motion) to increase with time, and to grow to macroscopic proportions.

At the level of a few individual photons (as opposed to stimulated emission in lasers, in which the presence of many photons is crucial), the propensity of two photons to bunch together was first observed experimentally by Hong, Ou and Mandel (HOM) \cite{hon87}. Specifically, when two separate photons are incident simultaneously upon a 50/50 beam splitter from opposite sides, the two output modes that emerge from the beam splitter each contain either two or zero photons, but never one photon each.
\begin{figure}[h]
\begin{center}
\includegraphics[width=1.8in]{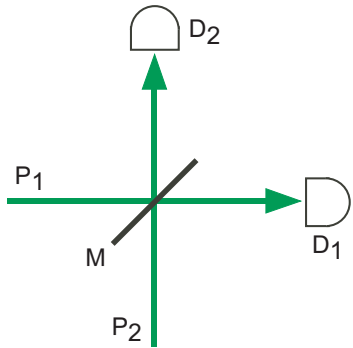}
\caption{Hong-Ou-Mandel interference at a beam splitter: Two single-photon wave packets ($P_1$ and $P_2$) impinge on a semi-transparent mirror ($M$) and the output wave packets are detected ($D_1$ and $D_2$).}
\end{center}
\end{figure}

Let us illustrate this effect in a simple case, which is illustrated in Fig. 1. (The complete theory is given in subsequent sections.) Two input modes (1 and 2) contain one photon each with the same center frequency, temporal width and polarization, which arrive at the beam splitter at the same time. The initial two-photon state is denoted by
\begin{equation}
|\Psi\rangle_{{\rm in}}=a_{1{\rm in}}^\d a_{2{\rm in}}^\d|{\rm vac}\rangle, \label{1.1}
\end{equation}
where $a_{1{\rm in}}^\d$ and $a_{2{\rm in}}^\d$ are creation operators for input modes 1 and 2, respectively. (These modes need not be monochromatic, but can be wave packets \cite{blow90,lou00}.) The relation between the input and output mode operators can be written as
\begin{eqnarray}
 a_{1{\rm in}}^\dagger&=&\tau a_{1{\rm out}}^\dagger - \rho a_{2{\rm out}}^\dagger, \nonumber\\
a_{2{\rm in}}^\dagger&=&\rho a_{1{\rm out}}^\dagger + \tau a_{2{\rm out}}^\dagger, \label{1.2}
\end{eqnarray}						
where the coefficients $\tau$ (transmissivity) and  $\rho$ (reflectivity) were assumed to be non-negative, without loss of generality, and satisfy the auxiliary equation $\tau^2 + \rho^2 = 1$. (Usually, one would give the output operators in terms of the input operators, as in the Heisenberg picture, but for our purposes the inverse transformations are more useful.) The input and output operators satisfy the commutation relations $[a_k,a_k^\d] = 1$ for k = 1 or 2. All other commutators are zero. By combining Eqs. (\ref{1.1}) and (\ref{1.2}), and using the fact that the vacuum is invariant under the action of the beam splitter, one finds that the output state is
\begin{eqnarray}
|\Psi\rangle_{{\rm out}}&=&
(\tau a_{1{\rm out}}^\dagger-\rho a_{2{\rm out}}^\dagger)
(\rho a_{1{\rm out}}^\dagger+\tau a_{2{\rm out}}^\dagger)
|{\rm vac}\rangle\nonumber\\
&=&\{\tau\rho[(a_{1{\rm out}}^\dagger)^2-(a_{2{\rm out}}^\dagger)^2]+(\tau^2-\rho^2)a_{1{\rm out}}^\dagger a_{2{\rm out}}^\dagger\}|{\rm vac}\rangle. \label{1.3}
\end{eqnarray}
In the case of exactly 50\% beam splitting, for which $\tau=\rho=1/\sqrt{2}$, Eq. (\ref{1.3}) reduces to
\begin{equation}
|\Psi\rangle_{{\rm out}} =
(|2\rangle|0\rangle-|0\rangle|2\rangle)/\sqrt{2}, \label{1.4}
\end{equation}
where $|n\>$ denotes an $n$-photon Fock state. (All the other modes are still vacuum modes.)
In this case the part of the state with one photon in each output port is zero. The physical interpretation of this cancelation is that there are two paths leading to the same final state component $|1\rangle|1\rangle$: Either both photons transmit through the beam splitter or both photons reflect. The probability amplitudes for these two events have opposite signs [$\tau^2$ and $-\rho^2$ in Eq. (\ref{1.3})], so they cancel. The resulting two-photon bunching is a basic property of nature and is at the heart of quantum information processing schemes using linear optics (beam splitters, polarizers, etc.) \cite{klm}.

Notice that this type of HOM interference does not depend on any relative phase between the input states: Unlike coherent states or classical light waves (whose interference properties at a beam splitter certainly depend on the relative phase of the complex amplitudes), Fock states do not possess any physically relevant phase on which interference could depend.

In the original setting \cite{hon87}, for the HOM effect to work, the photons must be in states that are identical in their polarization and spectral degrees of freedom (but their input states differ in their propagation directions). The question arises: Must two photons (or other bosons) have identical properties before they come together for this type of interference to occur? For example, assuming they have identical polarization states, must two photons initially be in identical energy (spectral) states in order to interfere?


In this paper, we show that the answer is negative: By using an active, frequency-shifting beam splitter, one can observe two-photon interference between photons of different color. The active process can be either reflection by a moving mirror (for small frequency shifts), or three-wave mixing in an optical crystal \cite{lou61,van04,alb04,rou04} or four-wave mixing in an optical fiber (for large frequency shifts) \cite{mck04,mck05,gna06,mec06}. The HOM effect is due to destructive interference in the output state, not the input state. We derive conditions on the forms of the input states that are required for high-visibility two-photon interference. These phenomena, in addition to demonstrating a fundamental property of light, will have applications in quantum information experiments involving photons of different color \cite{rod04,fer04,pfi04,mck08}.

\section{Passive Beam Splitters}

Henceforth, we will use a simplified notation: Instead of writing relations that define the input operators in terms of output operators, as in Eq. (\ref{1.2}), we will simply write $a_1^\dagger \mapsto \tau a_1^\dagger - \rho a_2^\dagger$ etc., to reflect how the input state changes to the output state in the most straightforward way. The operators $a_1$ and $a_2$ appearing on the right-hand side are thus meant to be output operators.

If the beamsplitter is a passive device that preserves the energy and spectrum of each photon separately, then the unitary transformation between the input and output channels can be written as
\begin{eqnarray}
a_1^\dagger(\omega) &\mapsto &\tau a_1^\dagger(\omega) - \rho a_2^\dagger(\omega), \nonumber \\
a_2^\dagger(\omega) &\mapsto &\rho a_1^\dagger(\omega) + \tau a_2^\dagger(\omega). \label{2.1}
\end{eqnarray}
Here we use creation operators $a_1^\dagger(\omega)$ for monochromatic modes with frequency $\omega$ impinging on one input port of the beam splitter and $a_2^\dagger(\omega)$ for modes impinging on the other input port. (We exploit a quasi one-dimensional picture of propagating single-photon wave packets. This picture is valid in the paraxial limit and when the radiation is sufficiently narrowband \cite{blow90,lou00}.) These operators satisfy the commutation relations $[a_k(\omega),a_k^\d(\omega')] = \delta(\omega - \omega')$ for $k = 1$ or 2. All other commutators are zero. Suppose now we have an input state containing two photons, one impinging on each input port, described by spectral amplitudes $\phi_1(\omega_1)$
and $\phi_2(\omega_2)$, respectively, which satisfy the normalization conditions
\begin{equation}
\int d\omega_k|\phi_k(\omega_k)|^2=1. \label{2.2}
\end{equation}
Then the input state is
\begin{equation}
|\Psi\>_{{\rm in}} = \int d\omega_1  \phi_1(\omega_1)
a_1^\dagger(\omega_1)
\int d\omega_2
\phi_2(\omega_2)
a_2^\dagger(\omega_2)|{\rm vac}\ra. \label{2.3}
\end{equation}
This input state is mapped by the beam-splitter transformation onto the output state
\begin{eqnarray}
|\Psi\>_{{\rm out}} = \int\int \hspace{-0.1in}&&d\omega_1 d\omega_2 \phi_1(\omega_1)\phi_2(\omega_2)[\tau\rho a_1^\dagger(\omega_1)a_1^\dagger(\omega_2) - \tau\rho a_2^\dagger(\omega_1)a_2^\dagger(\omega_2) \nonumber \\
&&+\ \tau^2a_1^\dagger(\omega_1)a_2^\dagger(\omega_2) - \rho^2 a_1^\dagger(\omega_2)a_2^\dagger(\omega_1)
]|{\rm vac}\ra. \label{2.4}
\end{eqnarray}
The last two terms can be made to cancel each other under the right conditions, and that complete destructive interference would correspond to HOM interference. The interference is perfect whenever $\tau^2 = \rho^2$ and
\begin{equation}
\phi_1(\omega_1)\phi_2(\omega_2)=\phi_1(\omega_2)\phi_2(\omega_1) \label{2.5}
\end{equation}
for {\em all} frequencies $\omega_1$ and $\omega_2$. This condition can be rewritten as
\begin{equation}
\phi_1(\omega_1)/\phi_2(\omega_1) = \phi_1(\omega_2)/\phi_2(\omega_2). \label{2.6}
\end{equation}
The left-hand and right-hand sides of Eq. (\ref{2.6}) are functions of different variables, implying that they cannot actually depend on those variables. Hence,
\begin{equation}
\phi_1(\omega_1)=C\phi_2(\omega_2), \label{2.7}
\end{equation}
where $C$ is a constant. Normalization of the quantum states gives $|C|=1$, so $C$ is just the phase factor $\exp(i\theta)$ for some phase $\theta$.\ Equivalently, in the time domain, the Fourier transforms of the spectral density functions must satisfy
\begin{equation}
\tilde{\phi}_1(t)=C\tilde{\phi}_2(t) \label{2.8}
\end{equation}
for destructive interference.

From these conditions one may get the impression that it is important to have indistinguishable photons at the input. But that is a little misleading; what counts is indistinguishability at the output, as we will show. The same output state can be reached by two different paths, and it is the interference between the two paths that matters (as {\em always} in quantum mechanics).

\section{Active Beam Splitters}

\subsection{Moving semi-transparent mirror}

Consider a semi-transparent mirror moving to the left in one dimension. In the mirror frame, two channels with the same carrier frequency (color), and opposite propagation directions (right and left) impinge on the mirror [Fig. 2($a$)]. The input and output channels are related by Eqs. (\ref{2.1}). However, in the laboratory frame [Fig. 2($b$)], light initally propagating to the right will be blue-shifted upon reflection, whereas light initially propagating to the left will be red-shifted upon reflection (and transmitted light does not change frequency).
\begin{figure}[htbp]
\begin{center}
\includegraphics[width=2.4in]{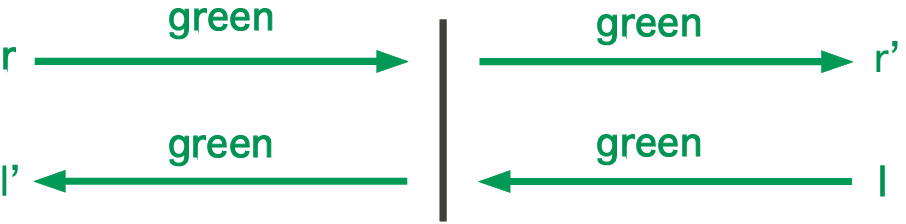} \hspace{0.4in} \includegraphics[width=2.4in]{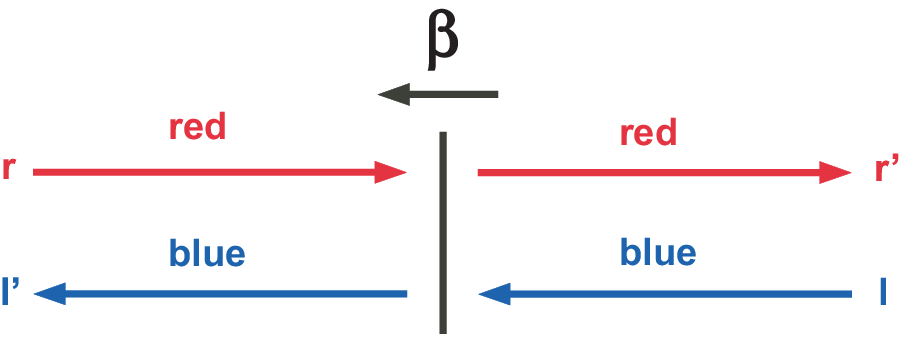}
\caption{Interaction of two waves ($r$ and $l$) at a semi-transparent mirror ($M$). In the mirror frame, the waves have the same frequency and opposite directions of propagation (right and left). In the laboratory frame, the waves have different frequencies and directions.}
\label{pump}
\end{center}
\end{figure}

It is convenient to label the right-propagating channel ``red'' and the left-propagating channel ``blue.'' (Of course, in order to shift actual red light to actual blue light and {\em vice versa}, the speed $v$ would have to be of order $c$.) The moving mirror transforms mode operators in the following way:
\ba
a^\dagger_R(\omega) &\mapsto &\tau a^\dagger_R(\omega) -
(\rho/\sqrt{\alpha})a^\dagger_B(\omega/\alpha), \nonumber \\
a^\dagger_B(\omega) &\mapsto &(\rho\sqrt{\alpha})a^\dagger_R(\alpha\omega) + \tau a^\dagger_B(\omega). \label{3.1.1}
\ea
The constant $\alpha$ is determined by the Doppler effect, which acts twice:
\begin{equation}
\alpha = (1 - \beta)/(1 + \beta) < 1, \label{3.1.2}
\end{equation}
where the normalized mirror speed is $\beta=v/c$. The factors of $\alpha$ in Eqs. (\ref{3.1.1}), which are absent from Eqs. (\ref{2.1}), are required to conserve the number of photons, because red modes in the frequency interval $d\omega_R$ are coupled to blue modes in the interval $d\omega_B = d\omega_R/\alpha$. Transformation (\ref{3.1.1}) is unitary, preserving the commutation relations for the transformed operators. For example,
\ba
&&[a_B(\omega),a^\dagger_B(\omega')] \mapsto
[\rho\sqrt{\alpha}a_R(\alpha\omega) + \tau a_B(\omega),\rho\sqrt{\alpha}a^\dagger_R(\alpha\omega') + \tau a^\dagger_B(\omega')] = \delta(\omega-\omega'), \nonumber \\
&&[a_B(\omega),a^\dagger_R(\alpha\omega')] \mapsto
[\rho\sqrt{\alpha}a_R(\alpha\omega) + \tau a_B(\omega),\tau a_R^\d(\alpha\omega') - (\rho/\sqrt{\alpha})a_B^\d(\omega')] = 0, \label{3.1.3}
\ea
where we used $\delta[\alpha(\omega-\omega')] = \delta(\omega-\omega')/ \alpha.$ Although this transformation preserves the number of photons, it does not preserve their energy, thus earning the right to be called active.

Now suppose we have one photon in each input mode, described by wave packets centered around red and blue frequencies, as the input state:
\begin{equation}
|\Psi\ra_{{\rm in}} = \int d{\omega_R}\phi_R(\omega_R)a^\dagger_R(\omega_R)
\int d{\omega_B} \phi_B(\omega_B)a^\dagger_B(\omega_B)|{\rm vac}\ra, \label{3.1.4}
\end{equation}
where the spectral amplitudes satisfy the normalization conditions (\ref{2.2}). This state is transformed by the moving mirror into
\ba
|\Psi\ra_{{\rm out}} = \int \int \hspace{-0.1in}&&d{\omega_R}d{\omega_B} \phi_R(\omega_R)\phi_B(\omega_B)[\tau a^\dagger_R(\omega_R) -
(\rho/\sqrt{\alpha})a^\dagger_B(\omega_R/\alpha)] \nonumber \\
&&\times [\rho\sqrt{\alpha}a^\dagger_R(\alpha\omega_B) + \tau a^\dagger_B(\omega_B)] |{\rm vac}\ra. \label{3.1.5}
\ea
The interesting effect is that there can be destructive interference in the output terms where one red and one blue photon emerge. There are two amplitudes for reaching that final state with opposite signs; the (necessary and sufficient) condition for exact destructive interference for all frequencies $\omega_R$ and $\omega_B$ is
\begin{equation}\label{23}
\tau^2\phi_R(\omega_R)\phi_B(\omega_B)=
\rho^2\phi_R(\alpha\omega_B)\phi_B(\omega_R/\alpha). \label{3.1.6}
\end{equation}
This condition implies that $\tau^2 = \rho^2$ and
\begin{equation}\label{24}
\phi_R(\omega_R)/
\phi_B(\omega_R/\alpha)=\phi_R(\alpha\omega_B)/\phi_B(\omega_B) \label{3.1.7}
\end{equation}
for all frequencies $\omega_R$ and $\omega_B$, which in turn implies that
\ba
\phi_R(\omega_R) &= &C\phi_B(\omega_R/\alpha), \nonumber \\
\phi_B(\omega_B) &= &C^{-1}\phi_R(\alpha\omega_B),  \label{3.1.8}
\ea
with $C$ a constant independent of $\omega_R$ and $\omega_B$. These two conditions are really the same. Normalization gives $|C|^2=1/\alpha$, so we can write $C=\exp(i\theta)/\sqrt{\alpha},$
with $\theta$ some constant phase.
In the time domain this condition gives
\begin{equation}
\tilde{\phi}_B(t):=\int d\omega_B \phi_B(\omega_B)\exp(i\omega_B t) = C^{-1}
\int d(\omega_R/\alpha) \phi_R(\omega_R)\exp(i \omega_R t/\alpha), \label{3.1.9} \end{equation}
which means that
\begin{equation}
\sqrt{\alpha}\exp(i\theta)\tilde{\phi}_B(t)=
 \tilde{\phi}_R( t/\alpha). \label{3.1.10}
\end{equation}
If the input wavepackets satisfy these relations, then the output state of the moving mirror will contain either two red photons moving to the right, or two blue photons moving to the left, but never one of each color.

\subsection{Frequency translation in fiber, simple version}

Now consider four-wave mixing in a fiber. Among the many processes that may occur, we are interested in frequency translation (Bragg scattering), in which two strong, classical pump waves ($p$ and $q$) convert ``signal'' ($s$) photons into ``idler'' ($r$) photons, or {\em vice versa} \cite{mck04,mck05,gna06,mec06}. The frequency-matching condition for this process, which is illustrated in Fig. 3, is
\be
\omega_q + \omega_r = \omega_s + \omega_p. \label{3.2.1}
\ee
It follows from Eq. (\ref{3.2.1}) that the difference between the signal and idler frequencies, $\omega_s - \omega_r = \Omega$, equals the difference between the pump frequencies, $\omega_q - \omega_p$. It is convenient to refer to the idler as the ``red'' sideband and the signal as the ``blue'' sideband. (Once again, we use these terms just to indicate two different colors.) If the pumps are monochromatic (CW), then each red frequency is coupled to one blue frequency. If the red frequencies all lie within the phase-matched bandwidth of the convertor, the red-to-blue conversion efficiency is frequency independent. If the pump powers and fiber length are chosen judiciously, the probability for frequency translation to occur for a given photon is 50\%, with the remaining 50\% probability assigned to the photon's frequency staying the same.
\begin{figure}[h]
\begin{center}
\includegraphics[width=2.4in]{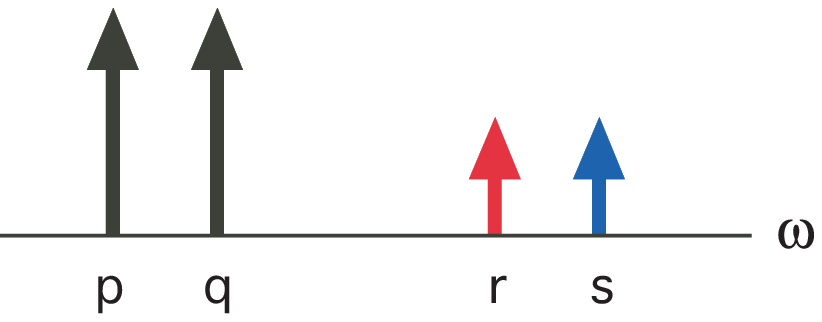} \hspace{0.4in} \includegraphics[width=2.4in]{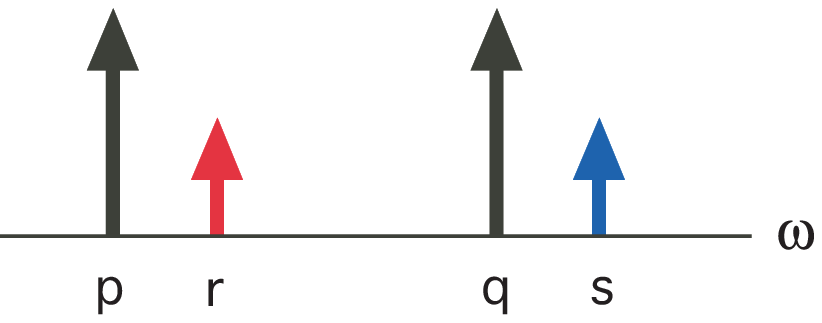}
\caption{Two classical pumps ($p$ and $q$) can drive a four-wave mixing process in which a single signal ($s$) photon is frequency-converted to a single idler ($r$) photon, or {\em vice versa}.}
\label{pump}
\end{center}
\end{figure}

The transformation between the red and blue modes is \cite{mck04,mck05}
\ba
a_R^\dagger(\omega_R) &\mapsto &\tau a_R^\dagger(\omega_R) - \rho a_B^\dagger(\omega_R+\Omega), \nonumber \\
a_B^\dagger(\omega_B) &\mapsto &\rho a_R^\dagger(\omega_B-\Omega) + \tau a_B^\dagger(\omega_B). \label{3.2.2}
\ea
(An equivalent transformation governs frequency up-conversion in a crystal \cite{lou61}.) This transformation is unitary. By following the steps between Eqs. (\ref{3.1.4}) and (\ref{3.1.6}), one finds that the HOM interference is perfect when
\begin{equation}
\tau^2\phi_R(\omega_R)\phi_B(\omega_B)=
\rho^2\phi_R(\omega_B-\Omega)\phi_B(\omega_R+\Omega). \label{3.2.3}
\end{equation}
Condition (\ref{3.2.3}) implies that $\tau^2 = \rho^2$ and
\ba
\phi_R(\omega_R) &= &C\phi_B(\omega_R+\Omega), \nonumber \\
\phi_B(\omega_B) &= &C^{-1}\phi_R(\omega_B-\Omega), \label{3.2.4}
\ea
with $C$ a constant, independent of $\omega_R$ and $\omega_B$. Once again, these two conditions are the same. Normalization gives $|C|=1$, and hence one can write $C=\exp(i\theta)$. In the time domain,
\begin{equation}
\exp(i\theta-i\Omega t)\tilde{\phi}_B(t) = \tilde{\phi}_R(t), \label{3.2.5}
\end{equation}
which means that the temporal shapes should be identical, with an overall frequency difference equal to $\Omega$. Equation (\ref{3.2.5}) provides necessary and sufficient conditions for frequency translation to display HOM interference in this simple case: Two input photons, one red and one blue, traversing a nonlinear fiber with the appropriate amount of frequency translation, lead to an output state which contains either two red photons or two blue photons, but never one of each color.

\subsection{Frequency translation in fiber, general version}

If one of the pumps is pulsed, or both pumps are pulsed, each frequency component of the signal is coupled to many frequency components of the idler and {\em vice versa} \cite{mck09}. In either case, the transformation over all frequencies is of the form
\begin{equation}
a^\dagger(\omega)\mapsto \int d\omega' G(\omega,\omega')a^\dagger(\omega'), \label{3.3.1}
\end{equation}
where the kernel $G$ satisfies the unitarity condition
\begin{equation}
\int d\omega' G(\omega,\omega') G^*(\omega'',\omega')=\delta(\omega-\omega''). \label{3.3.2}
\end{equation}
Let us split the frequency interval into two non-overlapping parts, ``red'' and ``blue,'' and let us correspondingly write $a(\omega)=a_R(\omega_R)$ whenever $\omega$ is a ``red'' frequency, and $a(\omega)=a_B(\omega_B)$ whenever $\omega$ is a ``blue'' frequency. Then we can rewrite the kernel in the block-matrix form
\begin{equation}
 \left[ \begin{array}{cc} G_{RR}(\omega_R,\omega'_R) &   G_{RB}(\omega_R,\omega'_B) \\
 G_{BR}(\omega_B,\omega'_R)&
 G_{BB}(\omega_B,\omega'_B) \end{array} \right], \label{3.3.3}
 \end{equation}
where the constituent (Green) kernels $G_{RR}$, $G_{RB}$, $G_{BR}$ and $G_{BB}$ describe red-red transmission, blue-red conversion, red-blue conversion and blue-blue transmission, respectively. The Green kernels satisfy the unitarity conditions
\begin{eqnarray} \label{3.3.4}
\int d\omega_R' G_{RR}(\omega_R,\omega'_R) G^*_{RR}(\omega_R'',\omega_R')
+
\int d\omega_B' G_{RB}(\omega_R,\omega'_B) G^*_{RB}(\omega_R'',\omega_B')
&=&\delta(\omega_R-\omega_R''), \nonumber \\
\int d\omega_R' G_{BR}(\omega_B,\omega'_R) G^*_{RR}(\omega_R'',\omega_R')
+
\int d\omega_B' G_{BB}(\omega_B,\omega'_B) G^*_{RB}(\omega_R'',\omega_B')
&=&0,
\end{eqnarray}
and similar conditions in which $R \leftrightarrow B$. In terms of these kernels, the transformations between the red and blue modes are
 \begin{eqnarray}
a_R^\dagger(\omega_R) &\mapsto &\int d\omega'_R G_{RR}(\omega_R,\omega'_R)a_R^\dagger(\omega'_R) +
\int d\omega'_B G_{RB}(\omega_R,\omega'_B)a_B^\dagger(\omega'_B), \nonumber \\
a_B^\dagger(\omega_B) &\mapsto &\int d\omega'_R G_{BR}(\omega_B,\omega'_R)a_R^\dagger(\omega'_R) +
\int d\omega'_B G_{BB}(\omega_B,\omega'_B)a_B^\dagger(\omega'_B). \label{3.3.5}
\end{eqnarray}
The interference of one red and one blue photon in the final state is perfectly destructive when
\begin{eqnarray}
\int\int \hspace{-0.1in} &&d \omega_R d\omega_B \phi_R(\omega_R) \phi_B(\omega_B) [G_{RR}(\omega_R,\omega_R')
G_{BB}(\omega_B,\omega_B') \nonumber \\
&&+\ G_{RB}(\omega_R,\omega_B') G_{BR}(\omega_B,\omega_R')] = 0 \label{3.3.6}
\end{eqnarray}
for all frequencies $\omega'_R$ and $\omega'_B$. By rearranging the terms in Eq. (\ref{3.3.6}), so that functions of $\omega_R'$ and $\omega_B'$ appear on different sides, one obtains the interference conditions
\ba
\int d\omega_R \phi_R(\omega_R)G_{RR}(\omega_R,\omega_R') &=
&C\int d\omega_B \phi_B(\omega_B)G_{BR}(\omega_B,\omega_R'), \nonumber
\\
\int d\omega_B \phi_B(\omega_B)G_{BB}(\omega_B,\omega_B') &= &-C^{-1}\int d\omega_R \phi_R(\omega_R)G_{RB}(\omega_R,\omega_B'), \label{3.3.7}
\ea
where $C$ is a constant, independent of frequency. The first condition states that the spectral amplitude of red output photons arising from red input photons be equal (up to a constant) to the spectral amplitude of red output photons arising from blue input photons. The second condition is similar, but for blue output photons.

In general, though, these conditions will not be fulfilled with given pumps in a given fiber. One needs to design the four-wave mixing process in such a way that with 50\% probability a red photon is either converted to a blue photon or remains a red photon with a possibly altered spectral wave function. The opposite transformations must hold for the incoming blue photon. The fiber dispersion, pump pulses, and the input red and blue wave packets must all be designed properly to ensure that the output red and blue wave packets are identical for either input channel. (This is work in progress.)

More precisely, if one uses the unitarity conditions (\ref{3.3.4}) to rewrite and combine Eqs. (\ref{3.3.7}), one obtains the interference conditions
\begin{eqnarray}\label{cond}
\phi_R(\omega_R)&=&C
\int d\omega_B
K(\omega_R,\omega_B)\phi_B(\omega_B), \nonumber \\
\phi_B(\omega_B)&=&C^{-1}
\int d\omega_R
K^*(\omega_R,\omega_B)\phi_R(\omega_R),\label{3.3.8}
\end{eqnarray}
where the HOM kernel $K(\omega_R,\omega_B)$ is
\begin{equation}
K(\omega_R,\omega_B)=\int d\omega'
[G^*_{RR}(\omega_R,\omega')G_{BR}(\omega_B,\omega') - G^*_{RB}(\omega_R,\omega')G_{BB}(\omega_B,\omega')]. \label{3.3.9}
\end{equation}
Recall that every hermitian kernel can be decomposed in terms of its eigenvalues, which are real, and a single set of eigenfunctions, which are orthonormal. In a similar way, every complex kernel has the Schmidt (singular-value) decomposition \cite{eke95,law00}
\begin{equation}
K(\omega_R,\omega_B) =\sum_n \sigma_n R_n(\omega_R) B_n^*(\omega_B), \label{3.3.10}
\end{equation}
where the coefficients (singular values) $\sigma_n$ are real and non-negative, and the mode functions $\{R_n(\omega_R)\}$ and $\{B_n(\omega_B)\}$ are self-orthonormal.
Subroutines are available, which determine the mode functions numerically. By substituting decomposition (\ref{3.3.10}) into Eqs. (\ref{3.3.8}), one finds that the interference conditions can be fulfilled only if there is at least one unit coefficient among the Schmidt coefficients. If there is exactly one unit coefficient, say $\sigma_1=1$, then one must have $\phi_R(\omega_R) = \exp(i\theta) R_1(\omega_R)$ and $\phi_B(\omega_B) = B_1(\omega_B)$ in order for complete interference to occur. If there are multiple unit coefficients, one can take superpositions of the corresponding modes. If the kernel possesses {\em only} unit coefficients, then one can choose any red spectral amplitude $\phi_R(\omega_R)$ and find a corresponding blue spectral amplitude. The simple cases of the preceding subsections are, in fact, such degenerate cases, whenever $\tau^2 = \rho^2$.
In the general case, which corresponds to pulsed pumps, one expects that only certain red and blue wavepackets, of finite bandwidth and duration, lead to perfect HOM interference. In all of these cases, $C$ must again be the phase factor $\exp(i\theta)$. Thus, the fiber and pump requirements for perfect HOM interference may be stated in a different and more succinct way: The HOM kernel must have at least one unit singular~value.

One can analyze the general case in an alternative way, by giving the Schmidt decompositions of the Green kernels directly, rather than the HOM kernel $K$. By using the results of \cite{was}, which are reviewed in the Appendix of this paper, one finds that
\begin{eqnarray}
G_{RR}(\omega_R,\omega_R')
&=&\sum_n \tau_n V_n(\omega_R) v_n^*(\omega'_R), \nonumber \\
G_{RB}(\omega_R,\omega_B')
&=&-\sum_n \rho_n V_n(\omega_R) w_n^*(\omega'_B), \nonumber \\
G_{BR}(\omega_B,\omega_R')
&=&\sum_n \rho_n W_n(\omega_B) v_n^*(\omega'_R), \nonumber \\
G_{BB}(\omega_B,\omega_B')
&=&\sum_n \tau_n W_n(\omega_B) w_n^*(\omega'_B), \label{3.3.13}
\end{eqnarray}
where the non-negative coefficients $\tau_n$ and $\rho_n$ satisfy the auxiliary equations $\tau_n^2 + \rho_n^2 = 1$, and the sets of mode functions $\{V_n\}$, $\{v_n\}$, $\{W_n\}$, and $\{w_n\}$ are self-orthonormal.
By combining Eqs. (\ref{3.3.9}) and (\ref{3.3.13}), one obtains the alternative decomposition
\be K(\omega_R,\omega_B) = 2\sum_n \rho_n\tau_n V_n^*(\omega_R) W_n(\omega_B). \label{3.3.14} \ee
It follows from Eqs. (\ref{3.3.10}) and (\ref{3.3.14}) that $\sigma_n = 2\rho_n\tau_n$, $R_n = V_n^*$ and $B_n = W_n^*$. Notice that $\sigma_n \le 1$, so the HOM condition $\sigma_n = 1$ represents a special (ideal) case.

The coefficients $\tau_n$ and $\rho_n$ can be interpreted as generalized beam-splitter coefficients, with $\tau_n$ playing the role of a transmission coefficient (not changing the color) and $\rho_n$ the corresponding reflection coefficient (changing the color), for each mode $n$. Specifically,
the Schmidt-mode expansions
\ba
\ a_R^\dagger(\omega)|_{\rm in} &= &\sum_n a_n^\dagger V_n(\omega),
\nonumber \\
\ a_B^\dagger(\omega)|_{\rm in} &= &\sum_n b_n^\dagger W_n(\omega),
\nonumber \\
\ a_R^\dagger(\omega)|_{\rm out} &= &\sum_n c_n^\dagger v_n(\omega),
\nonumber \\
\ a_B^\dagger(\omega)|_{\rm out} &= &\sum_n d_n^\dagger w_n(\omega), \label{3.3.15}
\ea
allow one to rewrite the continuous input and output operators for red and blue light in terms of the discrete operators $\{a_n,b_n,c_n,d_n\}$ \cite{blow90}.
In terms of these specially constructed discrete modes, the output operators are related to the input operators in the simple, pair-wise manner
\begin{eqnarray}
a_n^\d &= &\tau_nc_n^\d - \rho_nd_n^\d, \nonumber \\
b_n^\d &= &\rho_nc_n^\d + \tau_nd_n^\d. \label{3.3.16}
\end{eqnarray}
Equations (\ref{3.3.16}) are equivalent to the two-mode equations (\ref{1.2}). As expected, a unit $\sigma_n$ corresponds to a 50/50 beam-splitter transformation with $\tau_n=\rho_n=1/\sqrt{2}$.

Moreover, description (\ref{3.3.13}) allows us to quantify the effect of imperfections in frequency translation. That is, suppose that the fiber process we actually designed does not have any coefficient $\tau_n$ equal to $1/\sqrt{2}$. Then the best we can do is to use the mode that corresponds to the transmission coefficient that is closest to $1/\sqrt{2}$, say $\tau_1\approx 1/\sqrt{2}$. In fact, if we choose in that case $\phi_R(\omega_R)=V_1^*(\omega_R)$ and $\phi_B(\omega_B)=W_1^*(\omega_B)$, then the amplitude of the component of the output state containing one red and one blue photon is
\begin{equation}
\langle {\rm vac}|a_R(\omega_R)a_B(\omega_B)|\Psi\rangle_{\rm out} = (\tau_1^2-\rho_1^2) V_1^*(\omega_R)W_1^*(\omega_B). \label{3.3.17} \end{equation}
This equation generalizes the corresponding term in Eq. (\ref{1.3}).
The deviation from perfect two-photon interference can be quantified as the probability to find one red and one blue photon (at any red-blue frequency pair) in the output, and (\ref{3.3.17}) shows that this probability is $P_{RB}=(\tau_1^2-\rho_1^2)^2$.

\section{Conclusions}

In this paper, we showed that one can observe interference between two photons of different color. These photons arise independently from separate sources before they come together and interfere. The important point is that both photons end up in the same final spectral state (or mode), even though they began in two spectrally distinct states. It is not necessary for them to be identical initially, except in the obvious sense that they are both photons, which are fundamentally identical particles (if one takes a particle view of photons). For example, if one ``red'' and one ``blue'' photon enter the active interferometer, then either two red or two blue photons exit, but never one of each color.

Observing such bosonic bunching of photons, whose colors are different initially, requires reversible and symmetric inelastic scattering processes. Examples of such processes include three-wave mixing (wavelength down-conversion) in a crystal \cite{van04,alb04,rou04} and four-wave mixing (Bragg scattering) in a fiber \cite{gna06,mec06}. Both types of process can shift photon wavelengths by several hundred nanometers, with high photon-conversion efficiencies. Such two-photon interference should find applications in quantum information science, in schemes using entanglement between optical modes of different color.

\begin{acknowledgments}
MGR wishes to dedicate his contribution to this paper to the memory of Krzysztof Wodkiewicz --- a friend for nearly three decades. He will always value the positive influence that Krzysztof had as a friend, teacher and collaborator. This work was supported in part by NSF grant ECCS-0802109.
\end{acknowledgments}

\appendix

\section{Decomposition of a Unitary Transformation}

In this appendix, it is shown that a $2n$-mode unitary transformation is equivalent to $n$ 2-mode beam-splitter-like transformations. As an illustrative example, consider the effects of a rotator (circular phase-shifter) on a monochromatic wave \cite{shu62,mck06}. Let $A_+$ and $A_-$ be the components of the $2 \times 1$ amplitude vector $A$ relative to the circularly-polarized basis vectors $E_+ = [1,i]^t/2^{1/2}$ and $E_- = [1,-i]^t/2^{1/2}$, respectively, and let $\theta$ be one-half of the differential phase shift. Then the input and output vectors are related by matrix equation
\be A(\theta) = M(\theta)A(0), \label{a1} \ee
where the $2 \times 2$ transfer matrix
\be M(\theta) = \left[\begin{array}{cc} e^{i\theta} & 0 \\
0 & e^{-i\theta} \end{array}\right] \label{a2} \ee
is diagonal, with complex-conjugate eigenvalues. Now let $A_x$ and $A_y$ be the amplitude components relative to the linearly-polarized basis vectors $E_x = [1,0]^t$ and $E_y = [0,1]^t$, respectively. Then the associated transfer matrix
\be M(\theta) = \left[\begin{array}{cc} \cos\theta & \sin\theta \\
-\sin\theta & \cos\theta \end{array}\right] \label{a3}. \ee
What appears as phase shifts relative to one basis appears as rotation relative to another. The second transfer matrix also describes the effects of a beam splitter on two input waves ($\tau = \cos\theta$ and $\rho = \sin\theta$).

Now consider the $2n \times 2n$ unitary matrix
\be M = \left[\begin{array}{cc} A & B \\
C & D \end{array}\right], \label{a4} \ee
where the $n \times n$ matrices $A$, $B$, $C$ and $D$ describe signal-signal transmission, idler-signal conversion, signal-idler conversion and idler-idler transmission, respectively. Our derivation of the beam-splitter decomposition of a unitary transformation is similar to the derivation of the Bloch--Messiah decomposition of a squeezing transformation \cite{blo62,bra05,mck09}. The unitarity conditions are
\ba \left[\begin{array}{cc} I & 0 \\ 0 & I \end{array}\right] &= &\left[\begin{array}{cc} (AA^\d + BB^\d) & (AC^\d + BD^\d) \\ (CA^\d + DB^\d) & (CC^\d + DD^\d) \end{array}\right] \nonumber \\
&= &\left[\begin{array}{cc} (A^\d A + C^\d C) & (A^\d B + C^\d D) \\
(B^\d A + D^\d C) & (B^\d B + D^\d D) \end{array}\right]. \label{a5} \ea
The complex matrix $A$ has the singular-value decomposition (SVD)
\be A = U_aD_aV_a^\d, \label{a6} \ee
where $U_a$ and $V_a$ are unitary, and $D_a = \diag(\alpha_j)$. In the standard SVD, the $\alpha_j$ are non-negative. However, one can also allow them to be negative. Similar SVDs exist for the other transmission matrices, in which $D_b = \diag(\beta_j)$, $D_c = \diag(\gamma_j)$ and $D_d = \diag(\delta_j)$. The first diagonal condition in Eq. (\ref{a5}) implies that
\be U_aD_a^2U_a^\d + U_bD_b^2U_b^\d = I, \label{a7} \ee
from which it follows that $U_a = U_b$ and $\alpha_j^2 + \beta_j^2 = 1$. The other diagonal conditions imply that $U_c = U_d$, $V_a = V_c$, $V_b = V_d$, $\alpha_j^2 = \delta_j^2$ and $\beta_j^2 = \gamma_j^2$. The off-diagonal conditions all imply that $\delta_j = \alpha_j$ and $\gamma_j = -\beta_j$. Let $E_{1j}$ be the $j$th column of $U_a$, $E_{2j}$ be the $j$th column of $U_d$, $F_{1j}$ be the $j$th column of $V_c$ and $F_{2j}$ be the $j$th column of $V_b$ (so $E_{1j}$ is an eigenvector of $AA^\d$, $E_{2j}$ is an eigenvector of $DD^\d$, $F_{1j}$ is an eigenvector of $C^\d C$ and $F_{2j}$ is an eigenvector of $B^\d B$). Then these vectors form self-orthonormal sets and
\be M = \sum_j \left[\begin{array}{cc} E_{1j}\alpha_jF_{1j}^\d & E_{1j}\beta_jF_{2j}^\d \\
-E_{2j}\beta_jF_{1j}^\d & E_{2j}\alpha_jF_{2j}^\d \end{array}\right]. \label{a8} \ee
Equation (\ref{a8}) is the beam-splitter decomposition \cite{was}. $F_{1j}$ and $F_{2j}$ are the input eigenvectors, whereas $E_{1j}$ and $E_{2j}$ are the output eigenvectors. It follows from Eq. (\ref{a8}) that
\be M^\d = \sum_j \left[\begin{array}{cc} F_{1j}\alpha_jE_{1j}^\d & -F_{1j}\beta_jE_{2j}^\d \\
F_{2j}\beta_jE_{1j}^\d & F_{2j}\alpha_jE_{2j}^\d \end{array}\right]. \label{a9} \ee
Thus, if the forward transformation is determined by the parameters $(\alpha_j,\beta_j,-\beta_j,\alpha_j)$, the backward transformation is determined by the parameters $(\alpha_j,-\beta_j,\beta_j,\alpha_j)$, just like the two-mode transformation (\ref{a3}). Furthermore, the roles of the input and output eigenvectors are reversed. Equations (\ref{a8}) and (\ref{a9}) remain valid if one replaces the aforementioned real parameters by the complex parameters $(\alpha_j,\beta_j,-\beta_j^*,\alpha_j^*)$ and $(\alpha_j^*,-\beta_j,\beta_j^*,\alpha_j)$ in the forward and backward transformations, respectively, where $|\alpha_j|^2 + |\beta_j|^2 = 1$.

The Green kernels of the text are related to the constituent matrices of this appendix by the relations $G_{rr} = A^t$, $G_{rb} = C^t$, $G_{br} = B^t$ and $G_{rr} = D^t$, from which it follows that the HOM kernel $K = A^\d B - C^\d D$. By combining this result with decomposition (\ref{a8}), one finds that
\be K = 2\sum_j \alpha_j^*\beta_j F_{1j}F_{2j}^\d . \label{a10} \ee
Equation (\ref{a10}) is consistent with Eq. (\ref{3.3.14}), which defines the mode functions of the text ($\phi_r$ and $\phi_b$) indirectly. These mode functions are defined directly by the interference conditions (\ref{3.3.8}). In the matrix notation of this appendix, they are the eigenvectors of $KK^\d$ and $K^\d K$, respectively. By combining the formula for $K$ with conditions (\ref{a5}), one finds that
\ba KK^\d &= &4A^\d AC^\d C, \label{a11} \nonumber \\
K^\d K &= &4B^\d BD^\d D. \label{a12} \ea
$A^\d A$ and $C^\d C$ have the eigenvalues $|\alpha_j|^2$ and $|\beta_j|^2$, respectively, and the (common) eigenvectors $F_{1j}$, whereas $B^\d B$ and $D^\d D$ have the eigenvalues $|\beta_j|^2$ and $|\alpha_j|^2$, respectively, and the (common) eigenvectors $F_{2j}$. Hence, the mode functions $\phi_{rj} = F_{1j}$ and $\phi_{bj} = F_{2j}$. The associated singular values $\sigma_j = 2|\alpha_j\beta_j|$ are the square roots of the (common) eigenvalues of $KK^\d$ and $K^\d K$. These results are consistent with Eq. (\ref{3.3.10}).

\end{document}